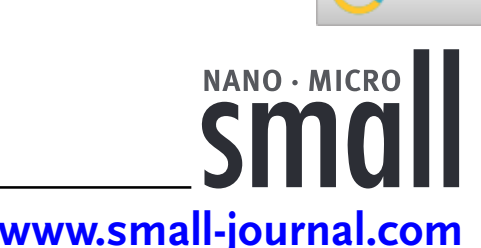

# DFT-Assisted Investigation of the Electric Field and Charge Density Distribution of Pristine and Defective 2D $WSe_2$ by Differential Phase Contrast Imaging

*Maja Groll, Julius Bürger, Ioannis Caltzidis, Klaus D. Jöns, Wolf Gero Schmidt, Uwe Gerstmann, and Jörg K. N. Lindner**

**Most properties of solid materials are defined by their internal electric field and charge density distributions which so far are difficult to measure with high spatial resolution. Especially for 2D materials, the atomic electric fields influence the optoelectronic properties. In this study, the atomic-scale electric field and charge density distribution of $WSe_2$ bi- and trilayers are revealed using an emerging microscopy technique, differential phase contrast (DPC) imaging in scanning transmission electron microscopy (STEM). For pristine material, a higher positive charge density located at the selenium atomic columns compared to the tungsten atomic columns is obtained and tentatively explained by a coherent scattering effect. Furthermore, the change in the electric field distribution induced by a missing selenium atomic column is investigated. A characteristic electric field distribution in the vicinity of the defect with locally reduced magnitudes compared to the pristine lattice is observed. This effect is accompanied by a considerable inward relaxation of the surrounding lattice, which according to first principles DFT calculation is fully compatible with a missing column of Se atoms. This shows that DPC imaging, as an electric field sensitive technique, provides additional and remarkable information to the otherwise only structural analysis obtained with conventional STEM imaging.**

## 1. Introduction

Transition metal dichalcogenides (TMD) are promising materials for next-generation optoelectronic devices due to their exciting physical properties.[1–6] 2D TMDs exhibit a variety of layer number-dependent properties, such as a tunable band gap,[7,8] as well as strong spin-orbit,[5,9,10] and light-matter interactions.[11,12] These layer number-dependent properties make 2D TMDs interesting for photonic and optoelectronic applications, e.g., for single photon emitters[13–17] and ultra-thin field effect transistors.[18,19] Tungsten diselenide ($WSe_2$) is such a semiconducting TMD. As common for TMDs, the bulk material of $WSe_2$ consists of stacked layers that are only bonded to each other via van der Waals interactions, while atoms within each individual layer are covalently bonded.[11,20] Reducing the number of layers, tungsten diselenide exhibits changing physical properties such as a transition from an indirect band gap of the bulk material to a direct band gap of the monolayer.[4,8] Point defects are omnipresent in bulk crystals for entropic reasons and are of crucial importance for numerous macroscopic properties. In a 2D material, crystallographic defects and lattice distortions may be assumed to influence macroscopic material behavior even more drastically due to the reduced dimensionality.[21] For example, defects at the edges of 2D $WSe_2$ flakes have been found to form single photon emitters.[22] Thus, the investigation of the crystallographic structure and electric field distribution of defect-free and defective 2D $WSe_2$ and other 2D TMDs is crucial for successful technological integration.

Experimentally, this always has been a challenging task. An emerging technique to achieve the visualization and quantification of local electric fields at sub-atomic resolution is differential phase contrast (DPC) imaging in the scanning transmission electron microscope (STEM), in short STEM-DPC. STEM-DPC has been applied to investigate the electronic structure of long-range electric,[23,24] piezoelectric,[25,26] and magnetic fields.[27] within a specimen. With state-of-the-art correction of lens aberrations which, in particular, includes the correction of the spherical aberration, STEM-DPC even allows us to visualize and quantify atomic electric fields with sub-atomic resolution.[28–32]

In STEM-DPC a convergent electron beam is scanned across the electron transparent specimen. The interaction of beam electrons with the electric fields present inside the specimen at each

M. Groll, J. Bürger, I. Caltzidis, K. D. Jöns, W. G. Schmidt, U. Gerstmann, J. K. N. Lindner
Department of Physics
University of Paderborn
Warburger Straße 100, 33098 Paderborn, Germany
E-mail: lindner@physik.upb.de

The ORCID identification number(s) for the author(s) of this article can be found under https://doi.org/10.1002/smll.202311635

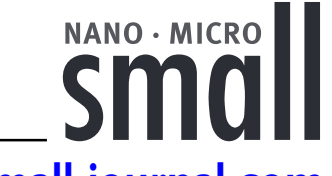

pixel of the scan causes a transfer of lateral momentum $\vec{p}_{\perp}$ perpendicular to the optical axis. For electric fields that are homogeneous across the probe diameter, the transfer of momentum leads to a measurable rigid shift of the electron beam intensity in the detector plane.[32,33] However, for atomic electric fields, which are usually inhomogeneous in direction and magnitude across the probe diameter, the beam-specimen interaction leads to a redistribution of intensity in the detector plane.[30,31,33,34] If the specimen is sufficiently thin and fulfills the (weak) phase object approximation, the transferred momentum is closely related to the center of mass (CoM) of the intensity distribution in the detection plane.[31,35,36] To measure the CoM in the detector plane, a position-sensitive detector such as a segmented is necessary. Segmented detectors determine the CoM by calculating the difference of the signals measured on opposing segments.[30,36,37] Quantum mechanical derivations correlate the transferred lateral momentum $\vec{p}_{\perp}$ measured by the shift of CoM with the lateral component of the atomic electric field $\vec{E}_{\perp}$ at the scan position $\vec{R}$ convolved with the probe intensity.[32,33] With the paraxial approximation the electric field can therefore be related to the transferred momentum by [38]

$$\vec{E}_{\perp}\left(\vec{R}\right) = \frac{\vec{p}_{\perp} m_{\text{rel}} v_{\text{rel}}^{2}}{p_0 e t} \tag{1}$$

Since the incident beam electrons with the elementary charge $e$ and the momentum $p_0$ are accelerated to high energies, the relativistic mass $m_{\text{rel}}$ and relativistic velocity $v_{\text{rel}}$ of the electrons must be considered in the calculations of the projected electric field. With the knowledge of the specimen thickness $t$ at the probe position $\vec{R}$ the lateral component of the projected (atomic) electric field $\vec{E}_{\perp}$ can be determined. In addition, the corresponding projected charge density can be obtained using Maxwell's equation $\rho_{\perp}(\vec{R}) = \epsilon_0 \nabla \cdot \vec{E}_{\perp}(\vec{R})$ where $\nabla$ is the 2D divergence and $\epsilon_0$ the vacuum permittivity.[31,32] It has been shown previously that quantification of STEM-DPC measurements requires the use of a few nanometers thin specimens.[33,39] Thus, 2D materials are well suited to further elaborate the STEM-DPC technique itself.

This work focuses on the employment of STEM-DPC to study the atomic electric field distribution of a pristine $WSe_2$ bilayer, whereby an unexpected distribution of the projected charge density over the 2 sublattices of the bilayer is observed. Furthermore, the electric field distribution of a Se vacancy column in a $WSe_2$ trilayer is investigated and the difference in field distribution to an intact region of the trilayer of $WSe_2$ is analyzed and discussed with the help of comparative DFT calculations. This demonstrates that the combination of the structural analysis joined with the electric field analysis enabled by DPC imaging can give a detailed insight into the electric field distribution of material associated with changes in the crystal structure.

## 2. Results and Discussion

$WSe_2$ flakes were mechanically exfoliated and transferred to holey silicon nitride TEM grids by a PDMS-based transfer process which allows for STEM investigations on free-standing 2D $WSe_2$ thin films. Schematics of the crystal structure of $WSe_2$ in [0001]- and [$11\bar{2}0$]-direction for the AA stacking configuration are shown in **Figure 1**a). Within this paper, we refer to the stacking sequence of the 2D $WSe_2$ flakes in the same way as described by He et al.[40] In this nomenclature, AA stacking indicates that metal atoms of the first layer are aligned with the metal atoms of the following layers, and the same holds for the chalcogen atoms. The investigation of a pristine bilayer of $WSe_2$ with a thickness of $t = (1.31 \pm 0.1)$ nm near [0001] zone-axis orientation is depicted in Figure 1b–e. The high-angle annular dark field (HAADF) as well as the DPC images are denoised using Gaussian blurring and non-rigid registration as described in the experimental section. In addition to the conducted measurements, Figure 1 shows the corresponding multislice image simulations of a $WSe_2$ bilayer in AA stacking configuration. More detailed information about the thickness determination by electron energy loss spectroscopy (EELS), image postprocessing, and simulations can be found in the experimental section and in the Supporting Information.

The HAADF image of the $WSe_2$ bilayer in Figure 1b) shows atomic columns with alternating high and dim intensities indicating an AA stacking of the 2 layers.[40–42] This alternating HAADF intensity (Z-contrast) is also visible in the simulated HAADF image on the right-hand side of Figure 1b). In $WSe_2$, this stacking configuration results in pure tungsten and pure selenium atomic columns. Based on the different HAADF intensities the position of selenium and tungsten atomic columns is determined in the following marked by green and gray dots, respectively.

Figure 1c) depicts the corresponding color-coded DPC image of the measured electric field distribution acquired with an eightfold segmented detector. The electric field direction at each pixel is indicated by the color according to the color wheel (inset) and the electric field magnitude is given by the intensity of the respective color.

Around the determined positions of tungsten and selenium atomic columns a color distribution roughly comparable to the inserted reference color wheel is visible in the experiment as well as in the simulation. Although a distortion of the color wheel and especially a reduced field magnitude is measured along the connecting axis between 2 atomic columns (example positions marked by gray arrows in Figure 1c), a radial electric field distribution is evident around all the atomic columns similar to those in references.[30,37]

The distortion of the color wheel is due to the small distance between tungsten and selenium atomic columns leading to an overlap of opposing electric field components and thus to a reduced momentum transfer. This effect is also visible within each hexagonal W-Se ring, where a local minimum of the measured projected electric field (marked by a white arrow in the measurement and in the simulation in Figure 1c) is observed near the geometric center due to the overlap of fields of 6 opposing atomic columns. These local field minima at highly symmetric points of the projected crystal structure due to an overlap of individual field distributions are described in the literature for different 2D materials with hexagonal crystal structures.[30,44–47]

In differential phase contrast imaging, no electric field is expected directly at atomic column positions.[30,37] The measured DPC image exhibits a non-zero electric field magnitude at almost all determined atomic column positions (see Figure 1c). Local field minima are not observed exactly at the atomic column positions determined using the HAADF image. However, local field minima are evident at positions slightly shifted compared to the atomic column positions in the HAADF image and are located at the bottom left of the position of the respective atomic column. This is particularly noticeable in the vicinity of the Se atomic columns (an example position is marked by an orange arrow in Figure 1c). These features might arise due to different lens aberrations influencing the measured DPC signal or specimen tilt during the measurement. The origin of these features is further investigated with the help of multislice image simulations shown in the Supporting Information. One of the resulting findings is that a small tilt of the specimen present during image acquisition leads to a change of the projected potential and thus to a distortion of the electric field distribution. Since DPC mainly uses the direct beam instead of incoherently scattered electrons, as is the case for HAADF image acquisition, DPC images are more sensitive to specimen tilt compared to HAADF images.[39,48] As specimen tilt during measurements is especially for thin materials difficult to adjust, Figure S7 (Supporting Information) shows the simulated HAADF and DPC images for a $WSe_2$ bilayer in dependence of increasing specimen tilt for up to 100 mrad. The specimen tilt is noticeable in the HAADF images and the charge density distribution by an elongation and merging of the detected intensity distribution of neighboring atomic columns in tilt direction. Based on these simulations, a local specimen tilt below 5° is estimated for the measurement shown in Figure 1. Another conclusion from simulations shown in the Supporting Information is that a shift of the local field minima with respect to the position of atomic nuclei cannot be explained by specimen tilt. This is presumably caused by residual 3-fold astigmatism as can be seen in Figure S8 in the Supporting Information.

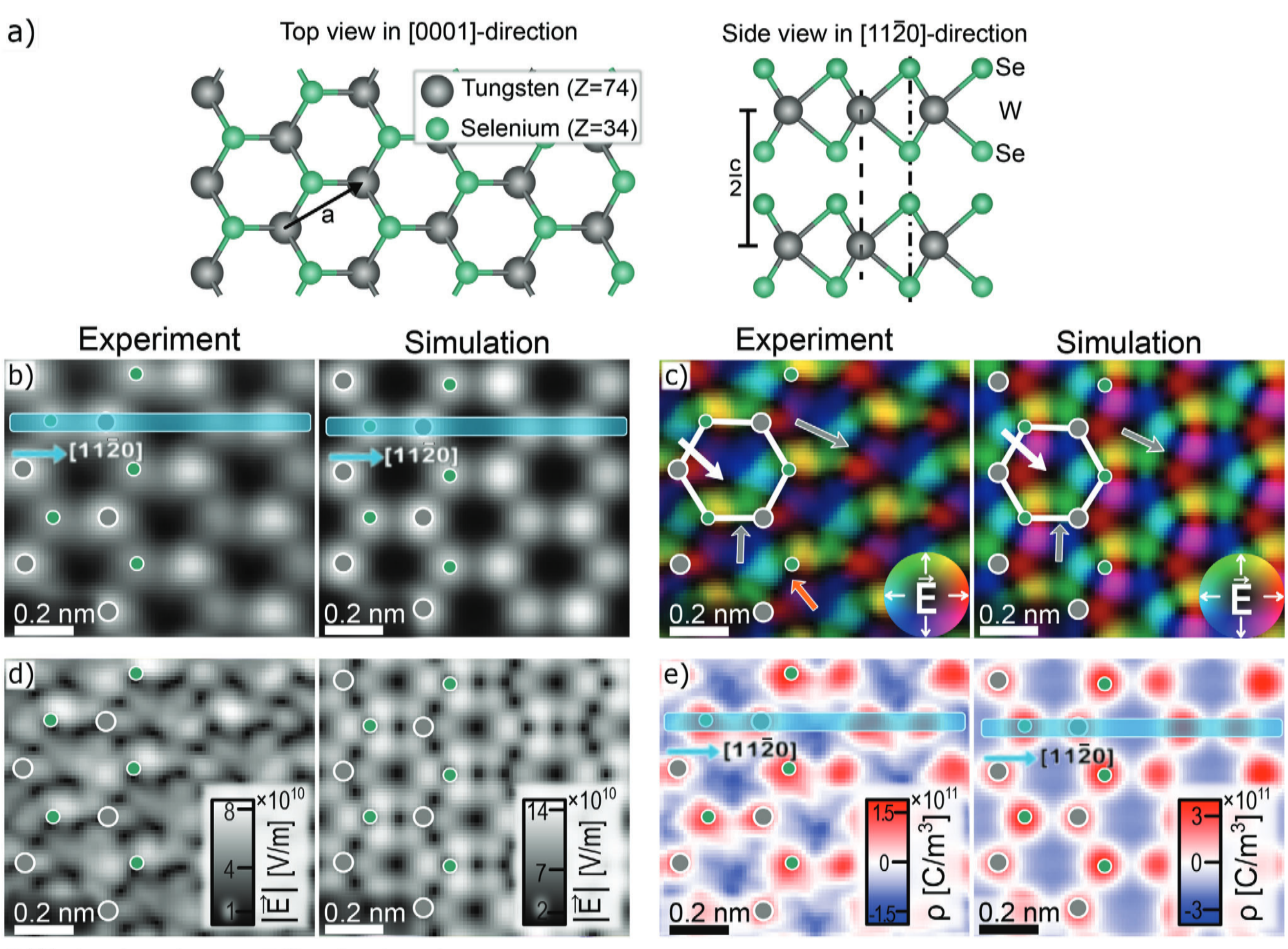

**Figure 1.** a) Sketch of the crystal structure of $WSe_2$ in [0001]- and [11$\bar{2}$0]-direction for AA stacking configuration.[40] The lattice constants indicated by the arrows are a = 0.329 nm and c = 1.298 nm.[43] For a bilayer of $WSe_2$, b–e displays a comparison between images acquired in the experiment (left) and images of the corresponding multislice image simulation (right). b) STEM-HAADF images of a bilayer $WSe_2$ in [0001] zone-axis orientation. c) Color-coded projected electric field maps and d) maps of the electric field magnitude. e) Corresponding charge density distributions. The position and type of the Se and W atomic columns are marked by green and gray dots, respectively.

Figure 1d) displays the electric field magnitude calculated using Equation (1). The measured electric field distribution is qualitatively in good agreement with the multislice image simulation (right-hand side of Figure 1d). Although the qualitative agreement is evident, the quantitative analysis reveals a discrepancy between the measured and the simulated electric field magnitudes. However, a direct quantitative comparison is difficult as the conducted multislice image simulations do not include any interatomic interactions and are only based on screened potentials of isolated atoms. In addition, the comparison is complicated by the fact that the measurable electric field magnitude can be reduced by non-optimal defocus and specimen tilt as well as residual lens aberrations.[39] The image post-processing and especially gaussian blurring can also slightly influence the measured electric field magnitude and therefore lead to differences between simulation and measurement. Differences between DPC measurements and analogous DPC simulations are already reported in the literature, but the origin is not clarified yet.[45,49]

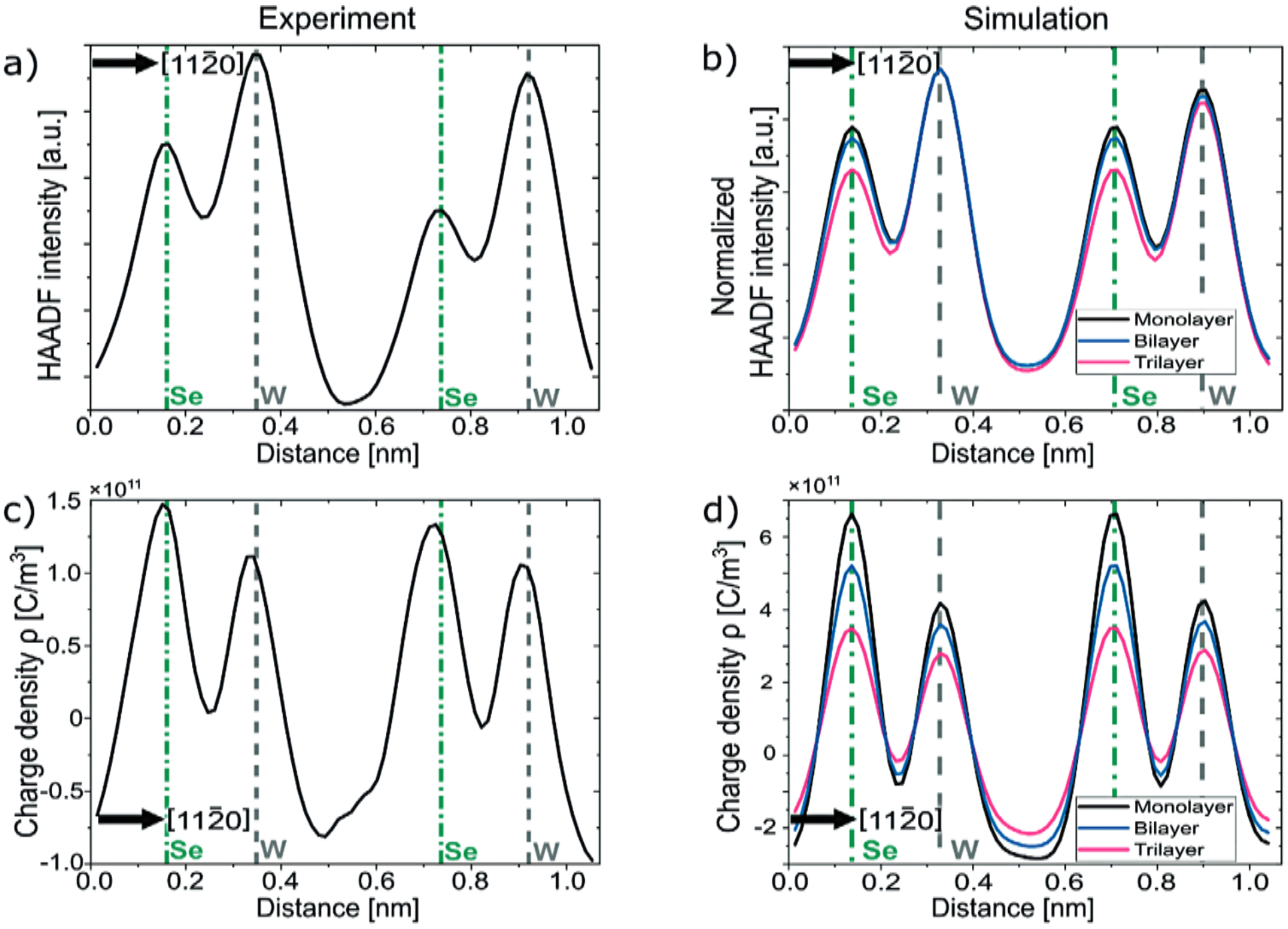

**Figure 2.** Line profiles of the HAADF intensity and charge density for the measurement and the simulations integrated across the width indicated by the blue marked regions in Figure 1b,e. The position and type of the atomic columns are marked by the green and gray dashed lines for selenium and tungsten atomic columns, respectively. a) and b) display the integrated HAADF intensity in [11$\bar{2}$0]-direction. Besides the HAADF intensity of a $WSe_2$ bilayer, b) also depicts the integrated HAADF intensity for simulated mono- and trilayer for the same region as exemplarily indicated in Figure 1b) for the bilayer. c) and d) show the integrated line profiles of the corresponding charge density for the area marked with a blue rectangle in Figure 1e). Again, d) also depicts the corresponding line profiles of the charge density for simulated mono- and trilayer integrated over the same area as indicated for the bilayer.

DPC measurements not only enable the investigation of atomic electric fields but also allow for the calculation of charge density distributions using Maxwell's equations. Figure 1e) shows the corresponding charge density distribution of the $WSe_2$ bilayer. As expected, the positive charge density is strongly localized at the position of the atomic columns and is dominated by the screened atomic nuclei. This is also reported in the literature for GaN and 2D molybdenum disulfide ($MoS_2$).[30,44,49,50] The negative charge density is delocalized within the hexagonal rings.[49]

For a further quantitative evaluation of the charge density distribution, **Figure 2** depicts integrated line profiles of the HAADF intensity and of the charge density from the measurement and the simulation, the latter for $WSe_2$ mono-, bi-, and trilayers. The areas of the integrated line profiles along the [11$\bar{2}$0]-direction are marked by blue boxes in Figure 1b,e and are integrated over 4 pixels for the measurement and simulation perpendicular to the [11$\bar{2}$0]-direction.

By analyzing the peak position of the integrated line profiles of the HAADF intensity, the position and the type of the atomic columns are determined and indicated by green and gray dashed lines in Figure 2 for selenium and tungsten atomic columns, respectively. Qualitatively, there is a good agreement between the experimental and simulated HAADF intensity profiles of the bilayer, showing the higher signal intensity at the W atomic column positions compared to Se columns. The simulations also demonstrate that the W to Se HAADF intensity ratio depends on the number of monolayers stacked on top of each other with the ratio increasing with an increasing number of stacked layers. One might want to exploit this to determine the thickness of the 2D $WSe_2$ flake inspected. However, a quantitative comparison between simulated and experimental intensity profiles like in Figure 2a) is hampered by the fact that recorded HAADF intensities largely depend on detector gain and contrast settings which are typically optimized individually for each image to obtain good image contrasts. Consequently, electron energy loss spectroscopy (EELS) was used to determine the thickness of the here investigated $WSe_2$ flake.

The atomic column positions from Figure 2a) are also inserted in the integrated line profiles of the charge density in Figure 2c,d. This demonstrates again that the local maxima of the positive charge density are not exactly at the determined positions of atomic columns due to the aforementioned 3-fold astigmatism.

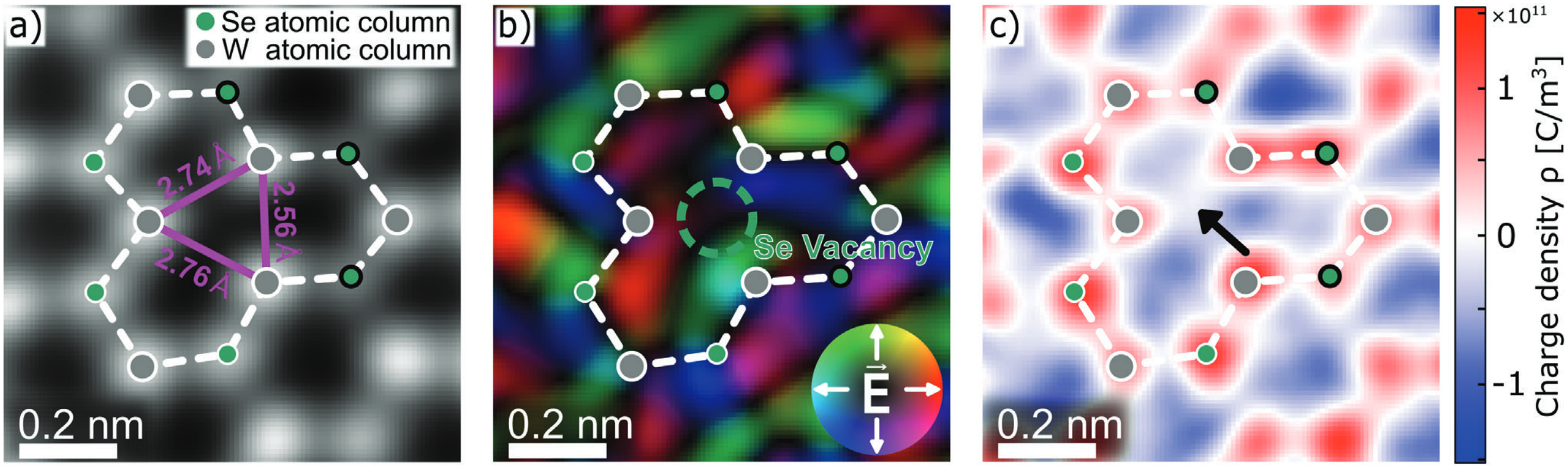

**Figure 3.** a) STEM-HAADF image of a point defect cluster in a trilayer of $WSe_2$ and b) the corresponding DPC image. c) Calculated charge density distribution. Selenium and tungsten atomic columns are marked by green and gray dots, respectively. The selenium atomic columns highlighted with bold black lines show a reduced intensity compared to other selenium atomic columns and might therefore exhibit a reduced number of atoms.

Unexpectedly, the integrated line profile of the experimentally obtained charge density in Figure 2c)reveals a higher positive charge density measured at selenium atomic columns than at tungsten atomic columns. In fact, the experimentally determined charge density averaged over all Se atomic columns in the scan area is higher by a factor of $1.48 \pm 0.29$ compared to that averaged over all W atomic columns within the same area.

This observation is surprising, as selenium has a lower atomic number ($Z_{Se} = 34$) than tungsten ($Z_W = 74$) and should therefore possess a lower positive charge density at the position of the atomic columns. Even though $WSe_2$ has twice as many selenium atoms within an atomic column compared to a tungsten atomic column, the projected atomic number in [0001] zone-axis orientation for AA stacking is lower for the selenium atomic columns than for the tungsten atomic columns. This unexpected ratio of the positive charge density of selenium and tungsten atomic columns is observed in different experimental DPC images of 2D $WSe_2$ flakes. The same behavior was also visible, although not directly reported, for monolayers and multilayers of molybdenum disulphide ($MoS_2$).[44,50] It is qualitatively confirmed in the conducted multislice image simulations for different numbers of layers shown in Figure 2d). The simulations for mono-, bi-, and trilayer show that the effect is most prominent when the flakes are as thin as one monolayer.

It is known that in general, quantitative analysis of DPC images must be conducted carefully as defocus and further residual lens aberrations influence the measurable electric field distribution.[39] However, these influences are not expected to drastically alter the ratio of positive charge densities at the atomic column positions of a binary compound. Since the effect is visible in both, experimentally determined and simulated charge density distributions, it can be ruled out that it is based on a charge redistribution in the 2D materials because such effects are not included in the simulations. In fact, according to our DFT calculations (Löwdin analysis) electron transfer between the 2 sublattices is small (9% of an electron per Se atom) and takes place from the Se atoms to the W sublattice. Instead, the observed higher positive charge density at selenium sites can presumably be explained by a coherent scattering effect which from atom to atom leads to an intensity redistribution inside the specimen. This effect will most likely depend on the kinetic energy of the electrons as well as the atomic scattering potentials. Thus, the obtained higher value for the charge density of the selenium atomic column is not necessarily the true projected charge density of selenium atomic columns. To quantify the influence of this effect an intensive investigation including the interaction of the electrons with the specimen potential and other specimen-related parameters is needed. Such a detailed description of this effect is beyond the scope of this article and will be given elsewhere.

Due to the low dimensionality, 2D materials are prone to defects that influence the electronic structure and thus the optoelectronic properties as well as the chemical reactivity. Information about the electronic structure in the vicinity of defects in such 2D materials is therefore essential. In the following, we show the structural changes and the changes in the electric field distribution induced by a defect in a $WSe_2$ trilayer. **Figure 3a**) depicts a STEM-HAADF image of the $WSe_2$ trilayer with a completely missing atomic column within the otherwise hexagonal structure. Here, the $WSe_2$ trilayer with a thickness of $t = (1.94 \pm 0.15)$ nm is again analyzed near [0001] zone-axis orientation, and again the selenium atomic columns and tungsten atomic columns are indicated by green and gray dots, respectively. Based on the HAADF intensity, an AA stacking has to be assumed, and thus the central pattern in Figure 3a) suggests the absence of a complete selenium atomic column (six selenium atoms). This is in agreement with the literature, as the selenium atoms have a lower displacement threshold energy than the tungsten atoms and are therefore more likely to be removed from lattice sites.[51]

The hexagonal atomic rings directly adjacent to the defect are slightly distorted which is illustrated by the white dashed line.

Those lattice distortions are reminiscent of atomic relaxation in the vicinity of defects as reported for different kinds of point defects.[52,53] Here, the lattice distortion is most prominently identified by the distance between the 3 tungsten atomic columns adjacent to the defect. The distances between the 3 atomic columns are measured and marked by the pink annotations in Figure 3a). With an average of ≈2.7 Å the distance is considerably reduced compared to the W-W distance in the ideal lattice as given by the lattice parameter of a = 3.29 Å.

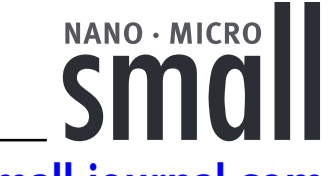

Three selenium atomic columns highlighted with thick black lines in the HAADF image in Figure 3a) exhibit lower HAADF intensities than other selenium atomic columns. From the literature, it is known that the formation of electron beam-induced defects in 2D TMD materials is a highly dynamic process that is driven by chalcogen vacancies and also includes the accumulation and diffusion of point defects.[47,54,55] We thus speculate that these marked atomic columns also have a reduced number of atoms or are hosting additional defects, i.e., lighter atoms like oxygen. Based on the HAADF peak intensities, which are reduced by ≈20 %–30 % compared to the mean HAADF intensity of selenium atomic columns in a pristine area, presumably up to 2 atoms (of 6) are missing within these atomic columns (further information can be found in Figure S4 in the Supporting Information). However, it should be noted that an additional Se vacancy is only one possibility, as, e.g., substitutional oxygen cannot be excluded.

In literature, it is often claimed that lighter atoms tend to fill up Se vacancies.[56] These substitutional atoms such as oxygen are less visible in HAADF images than tungsten or selenium atoms due to the detection principle of HAADF imaging. Instead, DPC imaging is more sensitive to light atoms as they introduce additional electric fields. The measured electric field map of this defect-including area is shown in the DPC image in Figure 3b). The surrounding hexagonal lattice in the vicinity of the defect reveals again some distorted but clearly visible color wheels around the atomic column positions and the inverted color wheels within the hexagonal atom rings. For the rings directly adjacent to the missing atomic column the central field minima are elongated, resulting in a reduced projected electric field at the position of the missing atomic column. No color wheel-like electric field distribution is visible at the position of the defect, supporting the interpretation of this defect as a completely missing atomic column with no substitutional atoms. Due to the missing atomic column, complete compensation of opposing electric fields of neighboring atomic columns is not possible, leading in sum to a projected electric field pointing toward the defect site. This is also visible in the map of the electric field magnitude shown in Figure S4b) (Supporting Information) as a field minimum is measurable at this position.

Based on the measured electric field distribution, the corresponding charge density distribution derived using Maxwell's first law is shown in Figure 3c). As described for the bilayer, the positive charge densities are localized at the positions of atomic columns, surrounded by negative charge density. Again, the charge density distribution reveals that some atomic columns exhibit a higher mean positive charge density than others. Based on the different HAADF intensities, selenium atomic columns show a 1.41 ± 0.26 times higher mean positive charge density compared to tungsten atomic columns.

Only for the 3 selenium atomic columns marked by the green dots with bold black contour lines the measured charge density is lower than the charge density at tungsten atomic columns. As mentioned above, this could be related to a reduced number of atoms within these 3 atomic columns. Furthermore, a slightly negative charge density is measured at the defect, but it is less negative compared to the surrounding charge density (indicated by the arrow in Figure 3c). To evaluate whether this change in the charge density distribution is induced by a substitutional atomic nucleus or induced by the neighboring atomic columns, multislice image simulations are performed with an increasing number of substitutional oxygen atoms in a missing selenium column of a $WSe_2$ trilayer. (Detailed information on the simulations can be found in the Supporting Information.) For up to 2 substitutional oxygen atoms, the measurable charge density at the defect site is still negative and only changes to a positive value for 3 oxygen atoms or more. To further quantify the change in measurable charge density induced by a certain number of substitutional oxygen atoms for a trilayer of $WSe_2$, relative charge densities are considered, i.e., the ratio of the charge density measured at the defect and the averaged charge density of the tungsten atomic columns are calculated. The averaged charge density of tungsten atomic columns is used as a reference because the transition metal atomic columns of TMDs are less prone to defects compared to the chalcogen atomic columns. Therefore, this value is reliable to be used as a reference value for the calculation and comparison with the simulation. As expected, the charge density ratios calculated from the simulations increase linearly with an increasing number of substitutional atoms (see Figure S6, Supporting Information). Comparing these ratios with the one calculated from the measurement it is concluded that presumably zero or a maximum of one substitutional oxygen atom fills up the defect. The uncertainty here is due to the fact that the simulated and measured values will depend on lens aberrations, the position, and nature of substitutional atoms within the defect, as well as the image blurring due to the electron beam.

The defect model of a missing Se atomic column in $WSe_2$ multilayers is also supported by first-principles calculations using density functional theory (DFT). For technical details please see the Methods section. By fully relaxing the atomic structure a 15 % inward relaxation of the 3 W dangling bond (db) atoms is obtained, resulting in W-W distances of ≈2.76 Å (see **Table 1**), which is in good agreement with experimental distances determined from the HAADF image (pink annotations in Figure 3a). Ref. [58] gives a systematic study on the structural relaxation caused by different kinds of defects in monolayers of various TMDs, i.e., $WSe_2$, which are in good agreement with our DFT calculations. As Ref. [58] only considers monolayers and neglects the interaction between stacked layers, we further investigate the structural relaxation of missing Se atomic columns for the case of bi- and trilayers.

**Table 1.** Calculated symmetry-conserving inward relaxation of the 3 central dangling-bond (db-) like W atoms around possible defects in 1L 2L, and 3L (trilayer) $WSe_2$ structures, as given by the resulting distance between 2 db-like W atoms. Notably, the size of the inward relaxation depends critically on the defect type (column of Se vacancies or $O_{Se}$ substitutional), but does not depend on the number of layers nor on spin-orbit coupling (SOC).[57] Even an additional Se vacancy right beside a missing Se atomic column (1L, 2+1 $vac_{Se}$) is only slightly distorting the triangle but leaves the average bond length within the central $W_3$ triangle almost unchanged. For the experiment, the average bond length is determined by the weighted average with the corresponding standard deviation.

| | PBE+D2 WW distance [Å] | PBE+D2 + SOC WW distance [Å] |
|---|---|---|
| $WSe_2$, *ideal* | 3.298 | 3.298 |
| 1 L, $2O_{Se}$ | 2.972 | 2.972 |
| 2 L, $4O_{Se}$ | 2.972 | 2.972 |
| 3 L, $6O_{Se}$ | 2.972 | 2.972 |
| 1 L, $2vac_{Se}$ | 2.763 | 2.766 |
| 2 L, $4vac_{Se}$ | 2.763 | 2.766 |
| 3 L, $6vac_{Se}$ | 2.763 | 2.767 |
| 1 L, $2+1vac_{Se}$ | 2.732 / 2.746 / 2.829<br>∅ = 2.769 | 2.736 / 2.742 / 2.827<br>∅ = 2.768 |
| exp. (this work) | 2.69 ± 0.03 | |

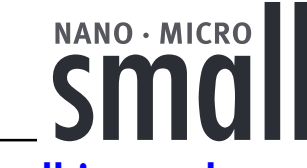

Clearly, the resulting distance between the central W atoms of 2.76 Å matches with the bond length of the nearest neighbors in metallic tungsten (2.74 Å [59]), strongly suggesting direct bonds, i.e., missing Se atoms. Based on the DFT total energy calculations, however, at least a completely O-decorated column can be clearly ruled out. The predicted inward relaxation for an O-decorated column is restricted to the 3 nearest-neighbor W atoms and with 5 % by far too small. Notably, the result that only undecorated Se-vacancy columns (i.e., $W_3$ cores) are compatible with the experiment, is very robust. The relaxation of the W dangling bond atoms does not depend on the number of $WSe_2$ layers (see Table 1). Also, spin-orbit coupling (SOC) has only a minor influence on the geometric properties and corrects the W-W distance of the db atoms only slightly, i.e., clearly below experimental accuracy of 0.001 nm. In addition, the 6 columns of neighboring Se atoms adjacent to the missing Se vacancy are shifted by 0.15 Å toward the W db atoms. As a result, the central $W_6Se_6$ unit (built up by 3 “honeycombs”) is somewhat separated from the rest of the 2D layer.

Comparing the atomic relaxation predicted by the DFT calculations with the structural state visible by the HAADF image, a completely missing selenium atomic column with no substitutional oxygen can be assumed.

As such defects alter the optoelectronic properties of the material, the exploration of resulting changes in the electric field distribution at the atomic level is of great interest. Investigations of the electric field distribution of point defects in 2D mono- or multilayers have so far only been described for molybdenum disulfide ($MoS_2$).[50] and hexagonal boron nitride (h-BN).[46] For monolayers of $MoS_2$, Calderon et al. observe a reduction in electric field magnitude for mono- and divacancies of missing sulfur atoms. There, a charge density distribution comparable to the one shown in Figure 3c) is reported for a missing S atomic column in a $MoS_2$ monolayer. For multilayers of h-BN, the electric field distribution in the vicinity of point defects is reported to be rather complex. Cretu et al. investigated a B vacancy in a h-BN multilayer and identified some regions of the defect-containing area with a locally reduced electric field magnitude compared to the electric field distribution in pristine h-BN and some regions in the vicinity of the defect with a locally enhanced electric field magnitude.

In a similar manner, we also analyze the electric field distribution at the defect site. More specifically, we consider the difference in the electric field $\Delta\vec{E} = \vec{E}_{def} - \vec{E}_{pris}$ between an almost pristine lattice $\vec{E}_{pris}$ and the defect-containing area $\vec{E}_{def}$, where the difference is calculated using a non-rigid registration [60] of atomic columns based on the HAADF images to register the defective and pristine areas. Here, both areas have the same number of pixels and depict an area of equal extension with ≈14 pairs of W-Se atomic columns. In addition, the defective area acts as the reference image to address the lattice distortion and the pristine image is registered to this reference image. The defect-related distortion of the lattice is therefore considered within this process. After registering, the difference in electric field magnitude is calculated. The result is depicted in **Figure 4a**). As indicated by the color bar, pixels with a blue color indicate a decrease in electric field magnitude, and pixels with a red color reveal an increased electric field magnitude compared to the pristine $WSe_2$ lattice.

This difference map clearly reveals a triangle-shaped region of reduced field magnitude at the defect site (highlighted by the dashed black triangle) with 3-fold symmetry. The tips of this triangle-like shape do not point toward the adjacent W atomic columns but are rotated by 60°. At the center of this region, almost no difference in electric field magnitude between the pristine and defective lattice is visible. However, this fits the expectations since no measurable electric field is to be expected either at the exact position of an atomic column or in a vacuum where opposing electric fields should compensate each other. More importantly, an increased electric field magnitude compared to the electric field magnitude of a pristine lattice is evident at the adjacent W atomic columns, as indicated by the black arrows in Figure 4a). This is in good agreement with the observation of Cretu et al. for a B vacancy in h-BN.[46]

To further clarify these observations, we conducted a multislice image simulation of a completely missing Se atomic column in a $WSe_2$ trilayer, where for the defective lattice the relaxed atomic positions obtained by the DFT calculations were used. Here again, no lens aberrations are considered and the focus is set to the entrance plane. Figure 4 b) shows the resulting difference in electric field magnitude $\Delta\vec{E} = \vec{E}_{def} - \vec{E}_{pris}$ between simulated DPC images of a defective and a pristine $WSe_2$ lattice. Also, here a characteristic triangular-shaped central feature with a decreased electric field is observed (black dashed triangle). Similar to the measurement, an enhanced electric field magnitude is localized in the vicinity of the 3 adjacent tungsten atomic columns (indicated by black arrows in Figure 4a).

With the help of DFT calculations, we can gain a more precise insight into the origin of this specific change in the electric field distribution, although a straight interpretation is difficult due to the different approaches of DPC imaging and DFT calculations. **Figure 5** shows the calculated spatial distribution of the defect-induced electron charge density of a $WSe_2$ monolayer in the side and top view (more details are given in the Methods section). In Figure 5a), the atomic structure with a missing selenium atomic column including the resulting atomic inward relaxation of the adjacent tungsten atomic columns and the corresponding electron charge density (red) is shown. Here, the atomic relaxation of the neighboring $W_6Se_6$ atoms discussed above becomes particularly obvious by the missing lines, i.e., missing covalent bonds, and by the db-like, crescent-shaped accumulation of electronic charge, pairwise bridging the 6 adjacent W atoms.

In contrast to the DFT results for the Se vacancy columns, the experimental HAADF, DPC as well and the calculated change in the electric field distributions in Figure 3 and Figure 4 do not reflect a perfect trigonal symmetry. This can be explained by the fact that even a single additional missing Se atom in this defective region breaks the symmetry completely if this vacancy is located in the same layer next to a missing Se column (see yellow arrow in Figure 5b). Interestingly, according to our DFT total energy calculations, such a '2+1' Se vacancy configuration is able to introduce strong symmetry-reducing changes in the surrounding electron charge, but keeps the triangular configuration of the central quasi-metallic $W_3$ core almost unchanged (see also Table 1). As a result, the distorted defect-induced electron charge (see Figure 5b) qualitatively resembles the experimentally determined induced change in electric field magnitude shown in Figure 4a). The other remaining features in the neighboring hexagonal atomic rings are due to polarization effects reflecting the DFT-derived inward relaxation of the central $W_6Se_6$ unit (cf. also Figure 4). Even though the electron charge distribution calculated by DFT resembles the triangular shape shown by the change in the electric field magnitude, one needs to carefully transfer this to one another as different approaches between DPC measurement and DFT calculations are used. However, the DFT calculations along with the DPC measurement provide a deeper insight into the induced changes caused by such a defect in a 2D material.

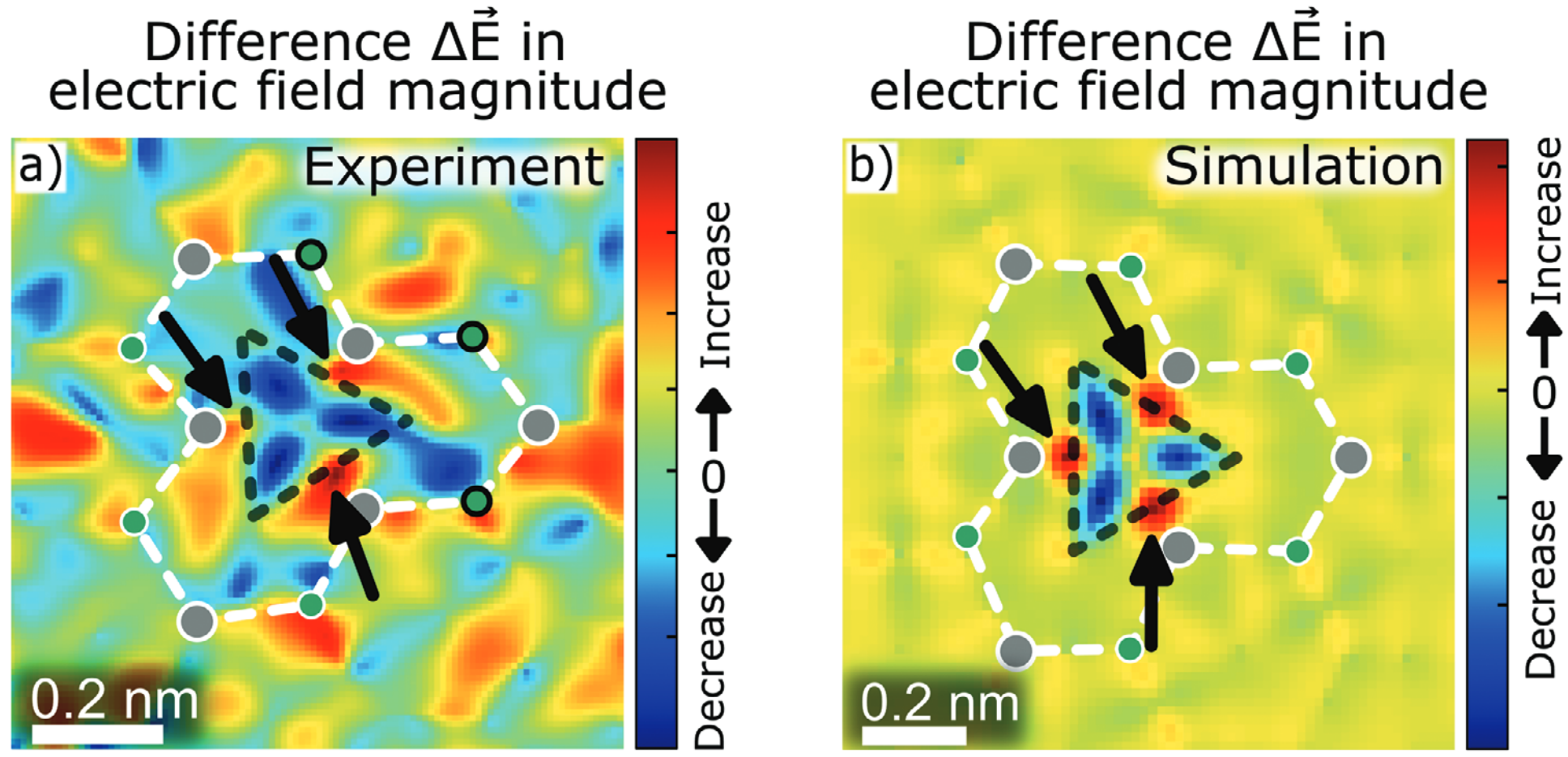

**Figure 4.** The difference in the electric field magnitude of the defective area compared to a pristine $WSe_2$ area of the same region for a) the measurement is shown in Figure 3b and 4b) a corresponding multislice image simulation of a missing selenium atomic column in a $WSe_2$ trilayer. For the simulation, the predicted atomic relaxation is considered.

In conclusion, the locally enhanced and reduced electric field distribution, visible in the simulation as well as in the measurement, is characteristic of the involved point defects, a missing column of Se atoms in a $WSe_2$ trilayer within AA stacking configuration. The observed local changes in the electric field and charge density distribution at defect sites can on the one hand change the optoelectronic behavior [9,61,62] and on the other hand act as adatom traps for diffusing atoms [46,50] and are therefore interesting for defect engineering . [62,63]

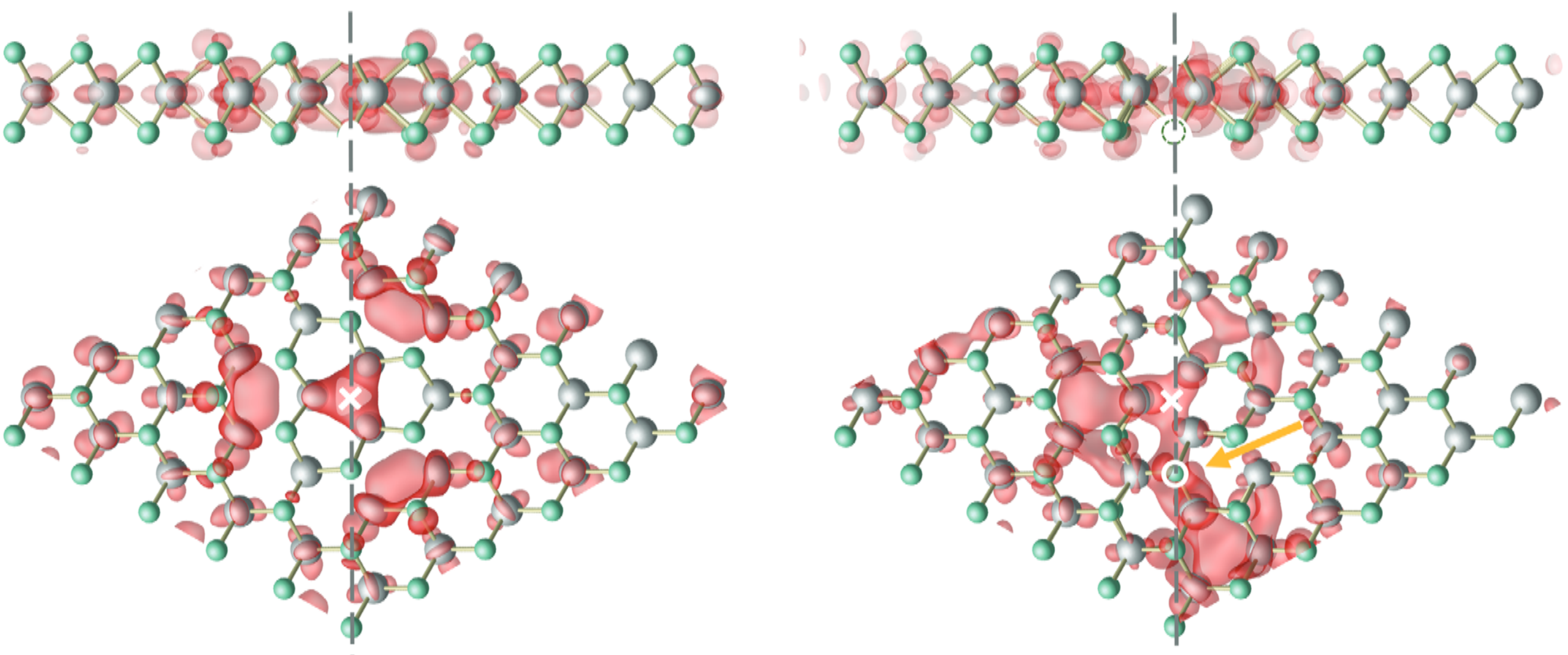

**Figure 5.** Atomic structure of the $WSe_2$ layer including defect structures and the electron density distribution of the related defect level (red). W and Se atoms in gray and green, respectively. The missing Se atomic columns are indicated by the vertical line and by the white cross (top view). a) Simple configuration with the Se vacancies restricted to a single column clearly showing trigonal symmetry. b) Defect configuration with an additional Se vacancy in one of the adjacent Se columns, indicated by the white circle and the yellow arrow.

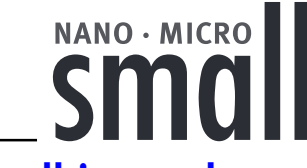

## 3. Conclusion

In this work, the atomic structure and – for the first time – the interatomic Coulomb field as well as the charge density distribution of 2D $WSe_2$ are investigated experimentally and theoretically, using STEM-HAADF and the emerging STEM-DPC (differential phase contrast) technique in combination with density functional theory (DFT). All investigated 2D flakes of $WSe_2$ exhibit the so-called AA stacking configuration, in which the metal atoms of stacked monolayers are aligned with the metal atoms of the adjacent monolayer(s), and the same holds for the chalcogenide atoms. The STEM-DPC measurements show outward pointing radial symmetric electric field distributions at the position of atomic columns and are in good qualitative agreement with the conducted multislice image simulations. The calculated charge density distribution of our measurement indicates that selenium atomic columns exhibit a higher positive charge density compared to tungsten atomic columns. This unexpected behavior is visible in measurements on a bilayer and trilayer of $WSe_2$ as well as in corresponding multislice image simulations and is ascribed to a coherent scattering effect. Because this effect seems to affect the interpretation of sub-atomic resolution STEM-DPC images of extremely thin specimens in general, further experimental and theoretical investigations are encouraged by our results.

Furthermore, the precise atomic configuration and electric field distribution of a point defect, namely a missing selenium atomic column in a $WSe_2$ trilayer, and the related symmetry-conserving relaxation of the surrounding lattice is analyzed and compared with DFT calculations. Deviations from perfect threefold symmetry are related to single Se vacancies in the vicinity of a missing atomic column. The difference in the electric field distribution of this defect and a pristine $WSe_2$ lattice is observed as a local triangular region of decreased electric field magnitude and a threefold symmetric region with enhanced electric field magnitude in the vicinity of the neighboring atomic columns. Qualitatively, the same behavior is obtained in multislice simulations of such a defect. All in all, STEM-DPC is shown to be a promising technique to not only reveal the electric field distribution of 2D materials in a pristine state but also is demonstrated to provide additional information about changes in the electric field caused by defects.

## 4. Experimental Section

*Fabrication and Transfer of 2D $WSe_2$ Flakes*: 2D $WSe_2$ flakes were mechanically exfoliated from a bulk crystal commercially available from 2Dsemiconductors Inc. (Scottsdale, AZ, USA) and transferred to PELCO holey silicon nitride TEM grids (Ted Pella, Inc., Redding, CA, USA) using a PDMS-based transfer process.[64] Further information on the transfer process is provided in the Supporting Information. The TEM grids are made of a 200 μm silicon substrate with a 0.5×0.5 mm window which is covered with a 200 nm thick holey silicon nitride membrane. The diameter of the holes is 200 nm, which allows TEM investigations of free-standing 2D TMD flakes.

*Characterization by Scanning Transmission Electron Microscopy*: High-resolution STEM investigations of 2D $WSe_2$ flakes are performed using a probe-side $C_s$-corrected JEOL JEM-ARM200F transmission electron microscope at an acceleration voltage of 80 kV. The microscope is equipped with a CEOS ASCOR $C_s$-corrector (CEOS GmbH, Heidelberg, Germany) which enables the correction of lens aberrations up to the fifth geometrical order including the correction of the spherical aberration ($C_s$). Additionally, a GATAN post-column GIF QUANTUM ER energy filter is used for thickness determination of the 2D $WSe_2$ flakes by electron energy loss spectroscopy (EELS).[65,66] Despite the presumed inaccuracy of EELS measurements on 2D materials, the local thickness using the log-ratio method is determined.[66–68] Based on previously performed photoluminescence measurements and due to the contrast of the $WSe_2$ flakes in visible light microscopy, the Malis model is chosen for the calculation of the mean free path as this fits best for the previously identified number of layers.[67,68] Thus, assuming an electron mean free path of 33.41 nm in $WSe_2$ at 80 kV, the thickness of the 2D flakes is determined via the log-ratio method using EELS measurements.

DPC measurements are conducted on an eight-fold segmented annular all-field (SAAF) detector consisting of 2 concentric rings subdivided into 4 quadrants each. The convergence semi-angle of the electron probe was fixed using a 40 μm condenser aperture and amounts to ≈(30.2 ± 0.8) mrad. The camera length was set to 12 cm such that the bright-field disk illuminates half of the outer 4 SAAF detector segments. All segments are individually fiber-optically coupled into photomultipliers connected to 32-bit AD converters, allowing for the independent detection of signals. This enables the acquisition of difference signals from opposing detector segments for the determination of the CoM of the bright-field disk. For the determination of the CoM in this work we use the outer 4 segments with an inner radial detection angle of $\beta$ = (18.2 ± 0.4) mrad and an outer detection angle of $\beta$ = (36.1 ± 0.4) mrad as this configuration approximates the real first momentum.[69] With the help of a previously conducted calibration, the calculated signal differences can be converted into the transferred momenta and allow for a quantitative analysis of the projected electric field. The calibration and the DPC image acquisition are done as described by Bürger et al.[39] Simultaneously with the DPC images, high-angle annular dark-field (HAADF) images are acquired using an annular detector with an inner polar angle of (54.2 ± 0.9) mrad and an outer angle of (191 ± 3) mrad for the camera length used.

*Post-Processing of Image Data*: DPC as well as HAADF images are post-processed using a drift correction and a Gaussian denoising algorithm which corresponds to a blurring with a Gaussian function with a FHWM between 55 and 90 pm (which equals a FWHM of ≈4 and 10 px). For the DPC investigations of a pristine $WSe_2$ bilayer, an additional HAADF signal-based non-rigid registration and subsequent averaging of various image segments is applied.[60,70] Furthermore, based on the Z contrast of HAADF images the atomic column positions and types are identified using a self-written algorithm that fits individual 2D Gaussian functions to the HAADF intensity distribution of individual atomic columns.[71] For the analysis of point defects, another specially developed program is used which enables the comparison of the electric field distribution and magnitude between a pristine and a defective lattice. In this algorithm, an image of the defective region is post-processed by the Gaussian denoising algorithm and a post-processed image of a nearby pristine lattice is used. In addition to Gaussian denoising, the image of the pristine lattice is post-processed by a rigid registration algorithm, a procedure that could not unequivocally be applied to images of defected areas. A pristine region near the defect region is chosen so that the possible influence of different defocus is reduced. In the second step, both images are overlayed using a non-rigid registration (NRR) algorithm based on an approach introduced by Thirion.[70] With the help of this algorithm, the position of the atomic columns of both images are aligned and the difference between the image of the pristine lattice and the image of the defect is calculated for each pixel separately. In order to take into account the possible lattice distortion in the defective area when applying the NRR algorithm to the 2 images, the image of the defective region is used as a reference image. Therefore, by correctly aligning the atomic columns of both images to each other and calculating the difference in electric field magnitude $\Delta \vec{E} = \vec{E}_{def} - \vec{E}_{pris}$ each pixel represents a difference $\Delta\vec{E}$ and thus allows to display a map of the difference in electric field magnitudes.

*DFT Modeling*: To model the defective $WSe_2$ layer structures, density-functional theory (DFT) using the Quantum ESPRESSO package was applied.[72,73] Single, bi-, and tri-layers of $WSe_2$ including up to 7 Se vacancies (or 6 oxygen atoms substituting Se atoms) are modeled using periodic boundary conditions (PBC) and $6 \times 6$ supercells, so that each $WSe_2$ layer contains 108 atoms, yielding system sizes up to 324 atoms. A shifted equidistant $3 \times 3$ $k$-point set is used to sample the Brillouin zone. Many-particle effects are taken into account using the semi-local Perdew-Burke-Ernzerhof (PBE) exchange-correlation functional,[74] whereby in all calculations the D3 dispersion correction [75] was used for a reasonable description of the van-der Waals interaction between $WSe_2$ layers. A plane-wave basis set with 800 eV kinetic energy cutoff and scalar relativistic gauge including projector augmented wave (GIPAW) pseudo-potentials was applied.[76] Besides scalar-relativistic calculations, fully relativistic two-component calculations are performed in order to investigate the influence of spin-orbit coupling (SOC).[57]

## Supporting Information

Supporting Information is available from the Wiley Online Library or from the author.

## Acknowledgements

M. G. gratefully acknowledges the financial support provided by a graduate scholarship from Paderborn University. W.G.S. and U.G. gratefully acknowledge the financial support by DFG (TRR 142/3-2023 – project number 231447078). The authors thank the Paderborn Center for Parallel Computing ($PC^2$) and the Höchstleistungs-Rechenzentrum Stuttgart (HLRS) for grants of high-performance computer time.

Open access funding enabled and organized by Projekt DEAL.

## Conflict of Interest

The authors declare no conflict of interest.

## Data Availability Statement

The data that support the findings of this study are available from the corresponding author upon reasonable request.

## Keywords